\RequirePackage{fix-cm}
\documentclass[smallextended]{svjour3}       
\smartqed  
\usepackage{graphicx}
\usepackage{natbib}
\setcitestyle{aysep={}}
\usepackage{graphicx,amsmath,amssymb}
\usepackage{url}
\usepackage{xfrac}
\usepackage{color}
\usepackage{hyperref}

\newcommand{\fref}[1]{Fig. \ref{#1}}
\newcommand{\sref}[1]{Section \ref{#1}}
\newcommand{\eref}[1]{(\ref{#1})}

\begin{document}

\title{Can the wave function in configuration space be replaced by single-particle wave functions in physical space?}

\titlerunning{Can the wave function in configuration}        

\author{Travis Norsen         \and
        Damiano Marian \and Xavier Oriols 
}


\institute{T. Norsen \at
              Smith College, Northampton, MA 01060, USA\\
              \email{tnorsen@smith.edu}           
              \and
              D. Marian (corresponding author)\at
              Dipartimento di Fisica dell'Universit\`a di Genova and INFN sezione di Genova, Via Dodecaneso 33, 16146 Genova, Italy \\
              Tel: +39-349-3681361\\
              \email{damiano.marian@ge.infn.it}
              \and
              X. Oriols \at
              Departament d'Enginyeria Electr\`onica, Universitat Aut\`onoma de Barcelona, Bellaterra 08193, Barcelona, Spain\\
              \email{xavier.oriols@uab.es}
}

\date{October 14, 2014}

\maketitle

\begin{abstract}
The ontology of Bohmian mechanics includes both the universal wave function (living in 3N-dimensional configuration space) and particles (living in ordinary 3-dimensional physical space).  Proposals for understanding the physical significance of the wave function in this theory have included the idea of regarding it as a physically-real field in its 3N-dimensional space, as well as the idea of regarding it as a law of nature.  Here we introduce and explore a third possibility in which the configuration space wave function is simply eliminated -- replaced by a set of single-particle pilot-wave fields living in ordinary physical space.  Such a re-formulation of the Bohmian pilot-wave theory can exactly reproduce the statistical predictions of ordinary quantum theory.  But this comes at the rather high ontological price of introducing an infinite network of interacting potential fields (living in 3-dimensional space) which influence the particles' motion through the pilot-wave fields.  We thus introduce an alternative approach which aims at achieving empirical adequacy (like that enjoyed by GRW type theories) with a more modest ontological complexity, and provide some preliminary evidence for optimism regarding the (once popular but prematurely-abandoned) program of trying to replace the (philosophically puzzling) configuration space wave function with a (totally unproblematic) set of fields in ordinary physical space.

\keywords{Local beables \and  Bohmian mechanics \and Conditional wave function \and Physical space \and Configuration space}
\end{abstract}

\tableofcontents

\section{Introduction}
\label{sec1}

Questions about how to understand 
the quantum mechanical wave function are, in this
current post-Bell renaissance period for quantum foundations,
increasingly popular and increasingly pressing.  
Some of these questions parallel historical disputes that have arisen
in the context of earlier theories proposing novel ontologies -- for
example, Newtonian gravitation
\citep{hesse,mcmullin,newton,janiak}
and electromagnetism
\citep{hunt,darrigol}.

But there are also some ways in which the debates about the status of the wave function are somewhat unprecedented.  For example, it is
unusual in the history of science for there to exist several so
radically different (but arguably empirically equivalent) theories,
such as we have with the Copenhagen, de Broglie - Bohm, spontaneous
collapse, many-worlds, and quantum Bayesian approaches to Quantum Mechanics (to name
just a few).  So there is not just one question about how to
understand the wave function, but instead many questions, about how to
understand the wave function in the context of each particular candidate
theory.

For the many-worlds theory, for example, the wave function is the only dynamical object in
the picture, so to whatever extent the theory is able to extract and
explain the elementary physical facts of everyday perceptual experience, it will
have to do so exclusively on the basis of the wave function.  For the
quantum Bayesian, the wave function represents not something physical
but instead something mental, subjective, informational; how such an
approach might account for elementary physical facts thus remains
obscure.  Whereas for the Bohmian, elementary physical facts are
accounted for in terms of something (namely, particles with definite
positions, and/or perhaps fields with definite configurations) that the
theory says exists \emph{in addition to} the wave function, which thus
in some sense plays a crucial (but background) role.  

We will focus here on this last possibility, the de Broglie - Bohm
pilot-wave theory, aka
Bohmian mechanics.  Our central concern here is developing the somewhat new idea that (despite some initial appearances to the contrary) Bohmian mechanics perhaps allows a uniquely appealing possible avenue for
addressing questions about the physical significance of the wave
function.  

We begin by explaining the initial appearance to the contrary.
As mentioned above, the wave function plays a somewhat
background role in the pilot-wave theory:  in the terminology of
\citet*{dgz}, for example, it is not part of the
\emph{primitive} ontology of the theory.  But it is still
\emph{there}.  Bell for example stressed that (compared to
some other interpretations in which the role of the wave function is
perhaps even less clear) it must really be taken seriously, as
corresponding to something physically real, in the context of the
pilot-wave theory:
\begin{quote}
``Note that in this [theory] the wave is supposed to be just as `real'
and `objective' as say the fields of classical Maxwell theory --
although its action on the particles ... is rather original.  \emph{No
  one can understand this theory until he is willing to think of
  $\psi$ as a real objective field rather than just a `probability
  amplitude'.  Even though it propagates not in 3-space but in
  3N-space}.'' \citep[p. 128, emphasis in original]{bell1}.
\end{quote}
We will focus in particular on this last point that for a general
Bohmian system of N particles (including in principle, indeed especially, the
universe as a whole) the wave function is, mathematically, a function
not on the usual 3-dimensional physical space, but is instead a
function on the 3N-dimensional configuration space.  But how can a
function on an abstract space like this possibly correspond to an objectively
real field?

Critics of the pilot-wave theory have often asked precisely this
question.  For example, N. David Mermin recently suggested that
advocates of the pilot-wave theory must (implausibly in his view)
give the ``3N-dimensional configuration space ... just as much
physical reality as the rest of us ascribe to ordinary
3-dimensional space.'' \citep{mermin}.  Less recently, 
Heisenberg made essentially the same criticism:  
\begin{quote}
``For [de Broglie and] Bohm, the particles are `objectively real'
structures, like the point masses of classical mechanics.  The waves
in configuration space also are objective real fields, like electric
fields....  [But] what does it mean to call waves in configuration space `real'?
This space is a very abstract space.  The word `real' goes back to the
Latin word `res', which means `thing'; but things are in the ordinary
3-dimensional space, not in an abstract configuration space.'' \citep{heisenberg}.\footnote{See also Chapter 8 of \citet{heisenberg2}}
\end{quote}
Proponents of the pilot-wave picture may find an easy rationalization
for dismissing such criticism, from the likes of Mermin and Heisenberg,
in the fact that their preferred alternatives are, to put it bluntly, non-sensical. 

But although he was not speaking of the pilot-wave theory
specifically, even the eminently-sensible Einstein seemed to share
this same concern, expressing grave doubts about the coherence of an
objectively real wave in configuration space.  He remarks, for
example, that
``Schr\"odinger's works are wonderful -- but even so one
nevertheless hardly comes closer to a real understanding.  The field
in a many-dimensional coordinate space does not smell like something
real.''  And similarly:  
``Schr\"odinger is, in the beginning, very captivating.  But the waves
in n-dimensional coordinate space are indigestible...''  (quoted in \citet{einstein})
It has been suggested that Einstein's negative reaction to Bohm's 1952
proposal -- which reaction almost everyone finds quite puzzling\footnote{In response to Einstein's remark that Bohm's
  way ``seems too cheap to me,'' for example, Max Born wrote that he
  thought ``this theory was quite in line with [Einstein's] own ideas,
  to interpret the quantum mechanical formulae in a simple,
  deterministic way...'' \citep[pp. 192-193]{born}} --  could be understood in this way:  although the theory
improved significantly on a number of intolerable aspects of the
Copenhagen approach, it simply retained this ``indigestible''
feature \citep{telb}.

Indeed, even Bohm himself seems to have found this aspect of his
theory (the idea of a physically real field living in an abstract configuration space) to be somewhat
indigestible:  
\begin{quote}
``...a serious problem confronts us when we extend the theory ... to
the treatment of more than one electron.  This difficulty arises in
the circumstance that, for this case, Schr\"odinger's equation (and
also Dirac's equation) do not describe a wave in ordinary
3-dimensional space, but instead they describe a wave in an
abstract 3N-dimensional space, where N is the number of particles.
While our theory can be extended formally in a logically consistent
way by introducing the concept of a wave in a 3N-dimensional space, it
is evident that this procedure is not really acceptable in a physical
theory, and should at least be regarded as an artifice that one uses
provisionally until one obtains a better theory in which everything is
expressed once more in ordinary 3-dimensional space.'' \citep[p. 117]{cnc}.
\end{quote}
Stepping back, there would seem to be three possible ways to
understand the wave function in the context of the Bohmian theory:

\begin{center}
{\bf{View 1: The universal wave function as a physically-real
  field in a physically-real configuration space}}
\end{center}

The first possibility is to simply bite the bullet and accept that the
3N-dimensional space in which the wave function lives must, if the
wave function is to correspond to a physically-objective (Maxwell-like)
field, be regarded as a (or perhaps the) physical space in its own
right.  For example, one could understand the theory as positing a
3-dimensional physical space (in which the particles move around) and,
in addition, a separate 3N-dimensional physical space (in which the wave
function lives).\footnote{It is not entirely clear, but Peter Holland may
  endorse this kind of view:  
``a complete and accurate account of the motions of
particles moving in accordance with the laws of quantum mechanics
\emph{must} be directly connected with multidimensional waves
dynamically evolving in configuration space.''  \citep[p. 321]{holland}.  Antony
Valentini has also made comments suggesting that the main innovation
of quantum mechanics is the need to accept the wave function as a new
kind of causal agent which physically affects particles despite its living
in a high-dimensional space. }  Such a view would evidently have to face several
obvious and pressing questions about the nature of the physical
process through which the wave function affects the particles.  To
avoid such questions, one might follow David Albert in proposing to
move the particles also into the high-dimensional space:  instead of N
points moving in 3-space, one might instead posit a single point
(sometimes referred to half-deprecatingly as ``the marvellous point'')
which moves under the influence of the wave function in 3N-space
\citep{albert}.  It is hardly clear that this is an improvement,
however, since one of the primary virtues of the original pilot-wave
theory (namely its ability to account for elementary physical facts
about the  everyday
macroscopic world in terms of the particles qua local
beables) is lost \citep{maudlin}.

In any case, many people -- such as Mermin, Heisenberg, Einstein, and
Bohm -- find themselves unable to 
stomach the non-local\footnote{We emphasize that the adjective ``non-local''
  has two different (but related) meanings. First, we use ``non-local''
  in the phrase ``non-local beable'' to denote an object (posited as
  physically real in some candidate theory) to stress that the object
  does not assign values to regions in 3-dimentional space (or 3+1
  spacetime).  And second, we use the word ``non-local'' to describe
  the special, faster-than-light, type of causal influence that
  Bell's theorem shows must exist \cite{bell1}.  In particular, we
  stress that despite proving the existence of non-locality (in the
  second sense) Bell's theorem does \emph{not} show that empirically
  viable theories must include beables that are non-local (in the first
  sense).  Indeed, one of the key points of our paper is a
  demonstration 
 that the non-locality required by Bell's theorem can actually
  be embedded in a theory of exclusively local beables. }
beables posited in this first view.

\begin{center}
{\bf{View 2: The universal wave function as a law}}
\end{center}

A second possible view of the wave function is to regard it as \emph{real},
but somehow not \emph{physical} in the sense of \emph{matter} or
\emph{stuff}.  In particular, it has been suggested that the wave
function should be thought of as playing a role, for the pilot-wave
theory, like the role that the Hamiltonian plays in the context of
classical mechanics.  The suggestion is thus that the wave
function should be regarded, not as a (Maxwell-like) field, but
instead as a \emph{law}.\footnote{For example, \citet*{law} write:
 ``We propose that the wave function belongs to an altogether
different category of existence than that of substantive physical
entities [--] that the wave function is a component of physical law
rather than of the reality described by the law.''} This is an interesting view that deserves
serious consideration.  However, its plausibility seems to rest on the
speculative idea that the wave function of the universe might be
static.  (It is somehow hard to swallow the idea of a law which
evolves non-trivially in time... with the evolution evidently being
governed by some further law.  Indeed, it is hardly clear that the
unusual notion of a law being governed by
another law is avoided even in the case of a static universal wave
function, for this too is supposed to be for example the solution of
the Wheeler - de Witt equation.)  The structural similarity between
Bohmian mechanics and electrodynamics also raises questions such as:
if the wave function should be understood as a law in Bohmian
mechanics, why not also interpret the electric and magnetic fields, in classical
electrodynamics, as laws?  This, of course, would be very strange and
nobody has ever suggested it.  But why not?  Evidently the answer is
that those fields,   
living on ordinary 3-dimensional physical space, \emph{can} be
unproblematically understood as material fields.  With no such option
apparently available for the quantum mechanical wave function, one
begins to get the feeling that View 2 is merely a convenient escape
hatch to avoid View 1.   

But perhaps after all there could be an option of interpreting the
quantum mechanical wave function in terms of material fields in
ordinary physical space.  We thus introduce:

\begin{center}
{\bf{View 3: The universal wave function as an abstract and indirect
      description of a set of single-particle wave functions in physical space}} 
\end{center}

In our proposed view, the wave function in a 3N-dimensional space might be
understood as a kind of indirect description of some more mundane sort
of physical stuff that lives in the ordinary, 3-dimensional physical
space and hence avoids the kinds of objections we have seen to
View 1.

The remainder of the paper will attempt to provide a case for
the viability of View 3 by showing how a concrete model (in which the
universal wave function is replaced by a set of one-particle wave
functions in physical space) can be devloped from Bohmian mechanics.

As a final point in this motivational overview, however, we note that View 3
has a kind of obvious plausibility to it, when we
consider historical examples of pre-quantum theories.  Indeed,
functions on high-dimensional, abstract (configuration and/or phase)
spaces are hardly unique to quantum mechanics.  For
example, the classical mechanics of N particles (usually
conceived in terms of N particles interacting through forces in
3-dimensional space) can be mathematically reformulated as a single
(marvellous?) configuration-space 
point $\vec{X} = \{X_1,X_2, ..., X_N\}$ moving under the influence of a
configuration-space potential $V(x_1, x_2, ..., x_N)$ according to
\begin{equation}
m_i \frac{d^2 X_i}{dt^2}  =  \left. - \nabla_i V \right|_{\vec{x} = \vec{X}(t)}
\label{eq-classmech}
\end{equation}  
where $m_i$ is the mass of the $i^\text{th}$ particle.\footnote{In this paper
we will use bold letters to indicate a point in the configuration
space, while a variable without arrow indicates a point in the
physical space.  Capital letters denote the actual positions of particles,
while lowercase letters denote generic positions.  In principle, positions in physical space have three
coordinates; however, for simplicity only, we will often consider a 1D
physical space. The symbol $\nabla_i$ accounts for the gradient in a
3D physical space or  $\nabla_i= \partial / \partial x_i$ in a 1D
physical space.}
Contemplating Equation \eqref{eq-classmech}, one might begin to worry
that classical mechanics
implies the existence of a kind of physical field on configuration
space, namely $V(\vec{x})$, whose gradient controls the motions of the
individual particles. 
Of course, the function $V$ being (at least if one contemplates
describing the entire universe)
time-independent, it is natural to think of it as having a law-like
(as opposed to matter-like) character.  

But the dynamics for these N classical particles can
also be re-expressed, as in the Hamilton-Jacobi formalism, in terms
of a non-trivially time-dependent function $S(x_1, x_2, ..., x_N,t)$ 
on the 3N-dimensional configuration space, evolving according to
\begin{equation}
\frac{\partial S}{\partial t} + \sum_i \frac{\left(\nabla_i
    S\right)^2}{2m_i} + V = 0
\label{eq-hj}
\end{equation}
and guiding the motion of the particles through
\begin{equation}
\frac{dX_i}{dt} = \frac{1}{m_i} \nabla_i S \big|_{\vec{x} = \vec{X}(t)}.
\end{equation}
To whatever extent one is
required to think of $S(\vec{x},t)$ as physically real (and to
whatever extent one dismisses the possibility that, if real, it might be
regarded as law-like rather than matter-like) the situation would
seem to be exactly parallel to the one in Bohmian mechanics:  the
theory involves a curious mathematical function on configuration space,
whose role is apparently to guide the particles in physical space.  

Why, then, were not people already worrying, in the 19th century, about
``indigestible'' physically-real fields on configuration space?
Evidently, the idea of interpreting (for example) Hamilton's principal
function, $S(x_1,x_2, ..., x_N,t)$, as an objective, real, Maxwell-like
field simply never even occured to anyone.  Instead it was always
regarded as obviously only an abstract mathematical reformulation, useful
perhaps for certain calculations, but not to be taken seriously as a
direct description of some piece of physical ontology.

The crucial point here is that this relaxed attitude is available, in
the case of $S(x_1,x_2, ..., x_N,t)$, precisely because there exist
\emph{also} the
alternative formulations, for example in terms of Newtonian forces, 
giving rise to the \emph{same} dynamical evolution for the N
particles, but in which everything in the mathematics can be readily
understood as corresponding to some physically real stuff in ordinary
3-dimensional space.  

And so the thought is:  maybe this is a clue to how we should try to
understand $\Psi(x_1,x_2,...,x_N,t)$ in quantum mechanics.  Maybe, that
is, the quantum $\Psi$ has the same status as the classical $S$ -- the
only difference being that, in the quantum case, we happened for
whatever reason to stumble first onto the abstract configuration-space
formulation of the theory and have not yet
managed to find the more directly physically interpretable 
alternative mathematical re-formulation in
terms of something like Newtonian forces, or (genuinely) Maxwell-like
fields (i.e., fields in 3-space rather than 3N-space), or perhaps some other,
wholly new, as-yet-uncontemplated type of local beables.  

Consideration of classical theories thus suggests that View 3 is
at least worth looking into.  In the rest of the paper we explain how
the Bohmian pilot-wave theory provides a particularly promising
starting point for this project via the
so-called \emph{conditional wave function} which, unlike the universal
wave function $\Psi$, can be understood as a field living on ordinary
3-dimensional physical space.  We begin, in Section \ref{sec2},
with an overview of Bohmian mechanics and, in particular, the
conditional wave function. In Section \ref{sec3} we lay out an important result for this work (based on an
idea first presented in \cite{telb}): the usual Bohmian particle trajectories can be reproduced in a theory
in which the configuration space wave function is replaced by
single-particle wave functions, if one introduces appropriate
time-dependent single-particle potentials. 
Section \ref{sec4} elaborates the concern that these novel potentials imply a kind of
infinite ontological complexity (perhaps even more problematic than
the questions about configuration space that the proposal is designed
to avoid) and then proposes a novel reformulation with a more modest
ontological complexity. This novel formulation is clearly preferable, on Occam's razor
type grounds, but also implies some disagreements with the usual
quantum mechanical predictions.  We thus present preliminary evidence, in
the form of numerical simulations, that the theory can
nevertheless be made empirically adequate (much like GRW type
theories, whose disagreements with ordinary quantum predictions are
confined to exotic situations not yet amenable to experimental test).  
Section \ref{sec5} then
summarizes the results and indicates some possible directions for
future work on the recommended ``View 3''.

\section{Bohmian Mechanics and the Conditional Wave Function}
\label{sec2}

We are primarily interested in the possibility, afforded especially by
Bohmian mechanics, of replacing the usual many-particle wave function
(in configuration space) by a set of one-particle wave functions (in
physical space).  To facilitate our discussion, we take the simplest
possible multi-particle system:  two spinless particles moving in one
spatial dimension.\footnote{The appropriate generalization for $N>2$
particles moving in 3 spatial dimensions is trivial; dealing with
systems of particles with spin in a fully general way 
probably requires working instead in terms of the Bohmian \emph{conditional density
matrices} defined by \citet*{cdm}, a possibility we set aside for
future work.}  

\subsection{Bohmian mechanics}

The description of this model system, according to Bohmian mechanics,
involves a wave function $\Psi(x_1,x_2,t)$ obeying the usual Schr\"odinger
equation
\begin{equation}
i \hbar \frac{\partial \Psi}{\partial t} = -\frac{\hbar^2}{2m_1}
\nabla_1^2 \Psi - \frac{\hbar^2}{2m_2}
\nabla_2^2 \Psi  + V(x_1,x_2,t) \Psi
\label{eq-sch}
\end{equation}
as well as positions $X_1(t)$ and $X_2(t)$ for the two particles.
These positions evolve according to the guidance formula
\begin{equation}
\frac{d X_i}{dt} = \left. v_i \right|_{\vec{x} = \vec{X}(t)} =
\frac{\hbar}{m_i} \, \left. \text{Im}
  \frac{\nabla_i \Psi}{\Psi} \right|_{\vec{x} = \vec{X}(t)}.
\label{eq-guidance}
\end{equation}
Note that the right hand side can also be understood as $v_i = j_i / \rho$,  evaluated at the actual configuration point $\vec{X}(t) = \{
X_1(t),X_2(t) \}$,  where $j_i$ is the standard \emph{probability current} associated with particle
$i$ and $\rho$ is the standard \emph{probability density}.  The
probability currents and density obey, as a consequence of
Equation \eqref{eq-sch}, the continuity equation:
\begin{equation}
\frac{\partial \rho}{\partial t} + \sum_i \nabla_i j_i = 0.
\label{eq-continuity}
\end{equation}
If we consider an ensemble of systems with the same wave function $\Psi$ but random
configurations $\vec{X}$ moving with configuration-space velocity $\vec{v}$, it is easy to see that the probability
distribution $P(\vec{X},t)$ over the ensemble should evolve according
to
\begin{equation}
\frac{\partial P}{\partial t} + \sum_i \nabla_i (P v_i) = 0 .
\end{equation}
Comparison with Equation \eqref{eq-continuity} shows that, since $v_i =
j_i/ \rho$,  $P=\rho$ is a solution.   Thus, if the initial
configurations $\vec{X}(0)$ in the ensemble are chosen with
distribution $P[\vec{X}(0) = \vec{x}] = \rho(\vec{x},0) =
\Psi^*(\vec{x},0) \Psi(\vec{x},0)$ --  the so-called \emph{quantum
equilibrium hypothesis} (QEH) -- it is then a consequence of
Equations \eqref{eq-sch} and \eqref{eq-guidance} jointly that 
\begin{equation}
P[\vec{X}(t) = \vec{x}] = \rho(\vec{x},t) = \Psi^*(\vec{x},t) \Psi(\vec{x},t)
\end{equation}
for all $t$. 
It is thus clear that Bohmian mechanics reproduces the statistical
predictions of ordinary quantum theory for position measurements, 
and hence also for any other type of measurement whose outcome is
ultimately registered in the position of some \emph{pointer}.  

\subsection{The conditional wave function}
\label{conditionalwave}

Ordinary quantum theory offers no way to define the wave function for
a single part of a larger quantum mechanical system, and it offers
only the notoriously vague and problematic measurement axioms to
describe how quantum mechanical systems interact with external,
separately postulated, classical systems such as measuring
instruments.  Bohmian mechanics, by contrast, treats the
whole universe in a uniform and fully quantum way.  It allows the
usual quantum measurement formalism (including for example the
collapse postulate, which of course has a dubious \emph{ad hoc} status
in the context of ordinary quantum theory) to be \emph{derived}, from the
fundamental dynamical postulates outlined above.
A key role in this analysis
(which was systematically presented by \citet*{bm,bm2})
is played by the \emph{conditional wave function}, which is the natural Bohmian
notion of the wave function for a sub-system. 

Here we will be particularly interested in treating each individual
particle as a sub-system of the universe.
The conditional wave function for particle $i$ is defined as the
universal wave function $\Psi$ evaluated at the actual
position(s) of the other particle(s).  For our toy two-particle
universe, the two conditional wave functions are:
\begin{equation}
\psi_1(x,t) = \Psi(x,x_2,t) \big|_{x_2 = X_2(t)}
\label{eq-conditional wave function1}
\end{equation}
and
\begin{equation}
\psi_2(x,t) = \Psi(x_1,x,t) \big|_{x_1 = X_1(t)}.
\label{eq-conditional wave function2}
\end{equation}
Each conditional wave function, because it depends only on the spatial coordinate associated
with the single-particle in question, can be regarded as a wave that
propagates in physical space.  The conditional wave functions are thus
a natural candidate for the status of centerpiece in a proposal to
re-interpret the configuration-space wave function $\Psi$ in terms of
many fields (and perhaps other sorts of local beables) living
exclusively in physical space.

Note that the guidance formula, Equation \eqref{eq-guidance}, may be re-written,
for each particle, in terms of its associated conditional wave function:
\begin{equation}
\frac{d X_i(t)}{dt} =
\frac{\hbar}{m_i} \, \left. \text{Im}
  \frac{\nabla_i \Psi}{\Psi} \right|_{\vec{x} = \vec{X}(t)} \equiv \frac{\hbar}{m_i} \, \text{Im}
\left. \frac{\nabla \psi_i}{\psi_i} \right|_{x = X_i(t)}. 
\label{eq-guidance2}
\end{equation}
The conditional wave function may thus be regarded as single-particle pilot-waves
(propagating in physical space) which guide the motions of the
affiliated particles.\footnote{Note that, as a result of Equations \eqref{eq-conditional wave function1} and \eqref{eq-conditional wave function2}, the conditional wave functions are not normalized in
the usual way.  It would be easy enough to adjust the definition to
yield normalized conditional wave functions;
but since any overall multiplicative factor cancels out anyway in
Equation \eqref{eq-guidance2}, we set aside this needless
complication.} 

\subsection{Conditional wave function evolution for non-entangled particles}
\label{non-entangled}

The conditional wave function
obeys the usual one-particle Schr\"odinger equation in the expected
type of situation in which the particle in question is suitably
isolated from its environment.  To see this, consider our simple
two-particle toy universe and suppose that the particles are initially
not entangled:
\begin{equation}
\Psi(x_1, x_2,0) = \alpha(x_1) \beta(x_2).
\end{equation}
Furthermore, suppose that the particles do not interact:
\begin{equation}
V(x_1, x_2, t) = V_1(x_1,t) + V_2(x_2,t).
\end{equation}
One can then show that 
\begin{equation}
\Psi(x_1,x_2,t) = \alpha(x_1,t) \beta(x_2,t)
\end{equation}
solves the Schr\"odinger equation, our
Equation \eqref{eq-sch} above, if $\alpha(x,t)$ satisfies the one-particle
Schr\"odinger equation
\begin{equation}
i \hbar \frac{\partial \alpha}{\partial t} = - \frac{\hbar^2}{2 m_1}
\frac{\partial^2 \alpha}{\partial x^2} + V_1(x,t) \alpha
\end{equation}
and similarly for $\beta(x,t)$.  

Thus, non-interacting particles which are not initially entangled will
remain unentangled for all time.  But the individual factors in the
factorizable (unentangled) wave function $\Psi$ are just (up to a
meaningless overall multiplicative constant; see footnote 8) the conditional wave functions
of the respective particles.  That is, with $\Psi(x_1,x_2,t) =
\alpha(x_1,t) \beta(x_2,t)$, we have 
\begin{equation}
\psi_1(x,t) = \Psi(x,x_2,t)|_{x_2 = X_2(t)} = \alpha(x,t) \beta(X_2(t),t)
\sim \alpha(x,t)
\end{equation}
and similarly $\psi_2(x,t) \sim \beta(x,t)$.  Thus, in the kind of
situation where ordinary quantum theory allows us to meaningfully talk
of the wave function of a single-particle, that wave function is
identical to the Bohmian conditional wave function and evolves in time according to the
expected, single-particle Schr\"odinger equation.

In general, though -- when there is
non-trivial interaction and hence entanglement among particles --  the conditional wave functions will evolve in an
unusual (non-linear and non-unitary) way.  To give a sense of the possible behavior and
also to provide a sense of how the measurement axioms of ordinary
quantum mechanics can instead be derived, from a careful analysis of
appropriate kinds of interactions, in Bohmian mechanics, we will
explain how the one-particle wave function associated with a quantum
system \emph{collapses} when there is a suitable interaction with
another system such as a \emph{measuring device}.  We illustrate this
in the following subsection with a simple
analytically-solvable toy model.

\subsection{Example of the non-unitary evolution of the conditional wave function}
\label{measurement}

Suppose that particle 1 is ``the quantum system to be measured'' and
particle 2 represents the center of mass coordinate of the macroscopic
pointer on a measuring device associated with quantum mechanical
operator $\hat{A}$.  Suppose the full system is initially in a product
state
\begin{equation}
\Psi(x_1,x_2,0^-) = \alpha(x_1) \beta(x_2)
\label{eq-initial}
\end{equation}
where $\alpha(x_1)$ is an arbitrary linear combination of $\hat{A}$ eigenstates
\begin{equation}
\alpha(x_1) = \sum_n c_n \alpha_n(x_1) \; \; \text{with} \; \; \hat{A}
\alpha_n(x_1) = a_n \alpha_n(x_1)
\end{equation}
and $\beta(x_2)$ is (say) a narrow gaussian packet centered at $x_2 =
0$ representing the measuring device in its \emph{ready} state. 

In a fully realistic description of a measurement, the ``quantum
system'' and the ``measuring device'' would need to interact in a way
that drives the wave function $\Psi$ into a set of
macroscopically-disjoint channels in the configuration space, with
each channel corresponding to a distinct perceivable possible outcome
of the measurement.  In the context of our simplified two-particle
toy-model, the overall idea can be adequately captured by positing, for
example, an impulsive interaction Hamiltonian like
\begin{equation}
\hat{H} = \lambda \, \delta(t) \hat{A} \hat{p}_{x_2}
\label{eq-interaction1}
\end{equation}
where $\hat{p}_{x_2}$ is the momentum operator for the pointer. 

In the special case that the ``quantum system'' happens
already to be in an
eigenstate $\alpha_m$ of $\hat{A}$, this interaction leaves the two-particle
system in the state $\Psi(x_1,x_2,0^+) = \alpha_m(x_1) \beta(x_2 -
\lambda a_m)$.   In light of the quantum equilibrium hypothesis, it
thus follows that the the actual position $X_2$ of the pointer will,
with unit probability, lie near the value $\lambda a_m$ indicating the 
pre-measurement (eigen)value $a_m$ of $\hat{A}$.  The pointer, in
short, displays the value normally associated with $\hat{A}$ in this
situation.  

In the general case, however, the interaction Hamiltonian
takes the state in Equation \eqref{eq-initial} to
\begin{equation}
\Psi(x_1, x_2, 0^+) = \sum_n c_n \alpha_n(x_1) \beta(x_2 - \lambda a_n)
\label{eq-final}
\end{equation}
in which the pointer is in an entangled superposition with the
system.  This final state reflects the notorious measurement
problem of ordinary quantum mechanics.  But for the Bohmian
pilot-wave theory, there is no problem:  the empirically observed
outcome is not to be found in the wave function, but instead in the
actual final pointer position $X_2(0^+)$.   Again in light of the
quantum equilibrium hypothesis, this will, with probability 
$|c_n|^2$, be near the value $\lambda a_n$ indicating that the outcome
of the measurement was $a_n$.\footnote{We assume here that the spacing, $|
\lambda(a_n - a_{n+1})|$, between adjacent possible pointer positions
is small compared to the width, $w$, of $\beta$.} 

Now consider how the conditional wave function for the ``quantum system'' evolves during the
measurement.  Prior to the interaction, when the joint two-particle
wave function is still given by Equation \eqref{eq-initial}, the conditional wave function
for particle 1 is
\begin{equation}
\psi_1(x,0^-) = \alpha(x) \beta(x_2) \big|_{x_2 = X_2(t)} \sim \sum_n c_n \alpha_n(x).
\end{equation}
This corresponds to the initial superposed wave function that would be attributed
to the ``quantum system'' in ordinary Quantum Mechanics.  But the post-interaction
$\Psi$, given by Equation \eqref{eq-final}, involves
disjoint channels of support in the configuration space.  The
final pointer position, $X_2(0^+)$, will randomly (depending on
initial conditions) end up in the support of \emph{just one} of these
channels.  That is, $\beta(x_2 - \lambda a_n)|_{x_2 = X_2(0^+)}$ will
(approximately) \emph{vanish} for all $n$ except the particular value, $n'$,
satisfying $X_2(0^+) \approx \lambda a_{n'}$, which corresponds to the
actual result of the measurement.  And so the
post-interaction conditional wave function for particle 1 will be
\begin{eqnarray}
\psi_1(x,0^+) &=& \sum_n c_n \alpha_n(x) \beta(x_2 - \lambda a_n)
\big|_{x_2 = X_2(0^+)} \nonumber \\
&\sim& \alpha_{n'}(x).
\end{eqnarray}
That is, the interaction causes the initially superposed wave function
for particle 1 to \emph{collapse} to the appropriate eigenfunction
(corresponding to the realized outcome of the measurement) even
though the wave function of the joint system evolves exclusively
unitarily according to the two-particle Schr\"odinger equation.
Bohmian mechanics thus \emph{explains}, from the point of view of a
theory in which all particles are treated in a fully uniform and fully
quantum way, how the wave functions of sub-systems may evolve exactly
as described (at the price of non-uniform treatment and additional 
\emph{ad hoc} postulates) in ordinary quantum mechanics.

\section{Wave equation for the single-particle wave functions}
\label{sec3} 

In the last Section we described the conditional wave function as
playing a certain role in the standard formulation of Bohmian
mechanics.  From this perspective, one crucial feature is that the conditional wave function
automatically evolves as sub-system wave functions should, according
to ordinary quantum mechanics.  (It is the fact that this happens
\emph{automatically} -- i.e., as a result of the fundamental dynamical
postulates, with no hand-waving and additional \emph{ad hoc}
``measurement axioms'' -- that is crucial and noteworthy here.)  But 
although the motion of the Bohmian particles can be expressed
exclusively in terms of their associated conditional wave functions, the conditional wave functions are not
usually thought of as having an independent existence.  They are,
after all, \emph{defined} in terms of the universal (configuration
space)  wave function,
and they do not (in general) evolve autonomously.  So they have a
status like, for example, the center of mass of a collection of
particles in classical mechanics:  they are a useful theoretical construct for
understanding certain features of the theory, but they do not have any
\emph{ontological} significance beyond that of the objects they are
defined in terms of.  

Our proposal -- a concrete implementation of ``View 3'' -- is to
reverse the ontological statuses usually assigned to the universal and
conditional wave functions.  That is, instead of regarding the
universal (configuration space) wave function $\Psi$ as ``physically
real'' (with the conditional wave functions being mere theoretical constructs), we propose
that the set of one-particle conditional wave functions can be invested with that primary
ontological status.  To explain this possibility, let us develop the
Schr\"odinger-type equations that can be understood as governing an
(almost) autonomous time-evolution for the conditional wave functions.  

\subsection{General Schr\"odinger-type equation for the single-particle wave functions}

A crucial point, underlying the behavior discussed in the
last Section, is that the conditional wave function (for, say, particle 1)
\begin{equation}
\psi_1(x,t) = \Psi\big(x,X_2(t),t\big)
\end{equation} 
depends on time in two ways:  through the Schr\"odinger
time-evolution of $\Psi$, and also through the time-evolution of
$X_2$.  
We may thus develop a Schr\"odinger-type equation for the
one-particle wave function of particle 1 as follows:
\begin{eqnarray}
i \hbar \frac{\partial}{\partial t} \psi_1(x,t) &=& i \hbar
\frac{\partial \Psi(x,x_2,t)}{\partial t} \Big|_{x_2 = X_2(t)} + i \hbar
\frac{d X_2}{dt} \frac{\partial \Psi(x,x_2,t)}{\partial x_2} \Big|_{x_2 =
  X_2(t)} \nonumber \\
&=& - \frac{\hbar^2}{2m_1} \frac{\partial^2 \psi_1(x,t)}{\partial x^2}
+ V\left[x,X_2(t),t\right] \psi_1(x,t) \nonumber \\
&& \; \; \; \; \; \;   + i \hbar \frac{d X_2}{dt}
\psi'_1(x,t) - \frac{\hbar^2}{2m_2} \psi''_1(x,t)
\end{eqnarray}
where we have defined
\begin{equation}
\psi'_1(x,t) = \frac{\partial \Psi(x,x_2,t)}{\partial x_2} \Big|_{x_2
  = X_2(t)}
\label{eq-psip}
\end{equation}
and
\begin{equation}
\psi''_1(x,t) = \frac{\partial^2 \Psi(x,x_2,t)}{\partial x_2^2}
\Big|_{x_2 = X_2(t)}. 
\label{eq-psipp}
\end{equation}  
The Schr\"odinger-type equation for $\psi_1$ can thus be re-written as 
\begin{equation}
i \hbar \frac{\partial \psi_1}{\partial t} = - \frac{\hbar^2}{2m_1}
\frac{\partial^2 \psi_1}{\partial x^2} + V_1^{\text{eff}}(x,t) \psi_1
\label{equation} 
\end{equation}
where 
\begin{equation}
V_1^{\text{eff}}(x,t) = V[x,X_2(t),t] + A_1(x,t) + B_1(x,t).
\label{eq-veff}
\end{equation}
This effective potential includes
the \emph{conditional potential} $V[x,X_2(t),t]$ (which is the usual
two-particle potential evaluated at the actual Bohmian location of the
other particle) plus some additional terms:
\begin{equation}
A_1(x,t) = i \hbar \frac{d X_2}{dt} \frac{\psi'_1(x,t)}{\psi_1(x,t)}
\label{A}
\end{equation}
and
\begin{equation}
B_1(x,t) = - \frac{\hbar^2}{2m_2} \frac{\psi''_1(x,t)}{\psi_1(x,t)}.
\label{B}
\end{equation}
It is important that the terms $A_1$ and $B_1$ in \eref{A}-\eref{B}
depend on $\psi_1(x,t)$ itself, making the whole equation non-linear.
(A different path
for the same deduction of the $A_1(x,t) $ and $B_1(x,t) $ terms can be
found in \citet{xoriolsPRL}.)  In addition, these terms can be complex,
so the time-evolution of the conditional wave function need not be unitary.
This explains how the conditional wave functions are able to exhibit wave-function
\emph{collapse}, as we saw in the previous Section.

It is interesting to note that the position
$X_2(t)$ of particle 2 has a direct influence on the time-evolution of
particle 1's conditional wave function, through the appearance of $dX_2/dt$ in the
term $A_1$.  And of course particle 1 likewise has a
direct influence on the evolution of particle 2's conditional wave function.  This is in
contrast to the usual formulation of Bohmian mechancs, in terms of the
configuration space wave function $\Psi$, in which the wave function
evolves completely independently of the particle positions.   Note
that this implies, for example, that in an ensemble of
identically-prepared systems (with identical initial wave functions,
but a distribution of initial particle positions) the conditional wave functions will evolve
\emph{differently} for the different members of the ensemble. 

The \emph{non-local} character of the dynamics can also be seen
here (we remind the reader about the two meanings of the adjective ``non-local'' mentioned in footnote 4). The dependence, for example, of $V_1^{\text{eff}}(x,t)$ on
$X_2(t)$ means that the pilot-wave field for particle 1 -- and hence
the motion of particle 1 itself -- can be influenced by interventions
which alter the trajectory $X_2(t)$ of the other (perhaps quite
distant) particle.  Our theory thus inherits the
dynamical non-locality of standard Bohmian mechanics, i.e., the sort
of non-locality that we know is required if one
wants to account for the empirically observed violations of Bell's
inequalities \citep{bell1}.  The interesting and important novelty here is that
our proposed theory is a (dynamically non-local) theory of exclusively
\emph{local beables}:  the particles and pilot-wave fields \emph{live}
in ordinary physical space, but the effective potentials
$V_i^{\text{eff}}$ which mediate their interactions imply
instantaneous actions at a distance.\footnote{On one hand, the
  potential $V[x,X_2(t),t]$ in \eref{eq-veff} produces correlations
  between the two particles $X_1(t)$ and $X_2(t)$. The dependence of
  $V[x,X_2(t),t]$ on $x$ and $X_2(t)$ imposes a restriction on the
  speed of such interaction. For  example, the \emph{retarded}
  electromagnetic potentials ensures that there is no superluminal
  electromagnetic influence between particles due to
  $V[x,X_2(t),t]$. On the other hand, such restriction on the
  \emph{speed} of the interaction between particles is not present in
  the new potentials $A_1(x,X_2(t),t)$ and $B_1(x,X_2(t),t)$ in
  \eref{eq-veff}. Thus, the particle $X_1(t)$ have an
  \textit{instantaneous} (non-local) interaction with $X_2(t)$ due to
  the potentials $A_1(x,X_2(t),t)$ and $B_1(x,X_2(t),t)$.  Of course,
  insofar as the usual Bohmian particle trajectories, and hence the
  usual statistical predictions of quantum mechanics, are reproduced,
  we know that this non-locality will not support superluminal
  communication.}
This is in contrast to standard
Bohmian mechanics, in which the non-locality is in some sense mediated
by the universal wave function, which of course lives in configuration
space (and is hence a \emph{non-local beable}, if it is a beable at
all).  

We stress, therefore, that the proposed theory (in which each
particle's motion is guided by an associated single-particle wave
function living in ordinary physical space) really does reproduce the
particle trajectories of standard Bohmian mechanics and hence the
exact statistical predictions of ordinary quantum theory.  In
particular, the dynamical non-locality that is manifest in the above
expressions for the (single-particle) effective potentials would allow
a (suitably generalized) theory of the type proposed here to account
for Bell inequality violations, quantum teleportation, and the various
other quantum phenomena which are sometimes erroneously thought to
\emph{require} a configuration space wave function.

\subsection{Defining the effective potential}
\label{secwaves}

By construction, $\psi_1$ obeying Equation \eref{equation} exactly reproduces
the usual Bohmian trajectory $X_1(t)$, and similarly for other particles.
Hence, a proper ensemble will exactly reproduce all 
statistical predictions of ordinary quantum theory -- provided the
one-particle effective potentials $V_i^{\text{eff}}(x,t)$ are defined
appropriately.  There is no difficulty with the conditional potential
terms, e.g., $V[x,X_2(t), t]$.  But our above definitions of $A_1$ and
$B_1$ involve $\psi'_1$ and $\psi''_1$ which are, in turn, defined in
terms of the configuration space wave function in Equations
\eref{eq-psip} and \eref{eq-psipp}.  So our proposed implementation of
``View 3''  -- formulating an empirically adequate quantum theory in
which the configuration space wave function is not present -- requires
finding a different way to define the single-particle effective potentials. Below, we discuss two different (but related) possibilities. 

One possibility for defining the terms $A_1$ and $B_1$ (without
reference to $\Psi$) can by found by adapting a proposal of
\cite{telb}.  For simplicity,
let us define the particle-1-associated potential fields $a(x,t)$,
$b(x,t)$, etc.,  as follows:
\begin{equation}
a(x,t) = \frac{\psi'_1(x,t)}{\psi_1(x,t)},
\label{fa}
\end{equation}
\begin{equation}
b(x,t) = \frac{\psi''_1(x,t)}{\psi_1(x,t)},
\label{fb}
\end{equation}
\begin{equation}
c(x,t) = \frac{\psi'''_1(x,t)}{\psi_1(x,t)},
\label{fc}
\end{equation}
and so on.
Then Equation \eqref{eq-veff} for 
the full effective potential which drives the evolution of particle
1's single-particle wave function $\psi_1$ can be re-written as 
\begin{equation}
V_1^{\text{eff}}(x,t) = V(x,X_2(t),t) + i \hbar \frac{dX_2}{dt} a(x,t)  -
\frac{\hbar^2}{2m_2} b(x,t).
\label{eq-v1eff2}
\end{equation}
We may then use the full configuration-space Schr\"odinger equation to
find out how $a(x,t)$ and $b(x,t)$ must evolve in order to exactly
reproduce the standard Bohmian trajectories.  The important thing here
is that the results can be written exclusively in terms of this
infinite network of potential fields.  For example,
the field $a$ should satisfy its
own partial differential equation of the form
\begin{eqnarray}
\frac{\partial a}{\partial t} &=& \frac{i\hbar}{2m_1} \left[
  \frac{\partial^2 a}{\partial x^2} + 2 \frac{\partial a}{\partial x}
  \frac{(\partial \psi_1 / \partial x)}{\psi_1} \right] + \nonumber \\
&& + \frac{i\hbar}{2m_2} \left[ c - ab \right] + \frac{dX_2}{dt}
\left[ b - a^2 \right] -\frac{i}{\hbar} \frac{\partial V}{\partial x_2} \Big|_{x_2 = X_2(t)}.
\label{eq-ade}
\end{eqnarray}
And similarly, $b$ will satisfy an evolution equation of the form
\begin{eqnarray}
\frac{\partial b}{\partial t} &=&  \frac{i\hbar}{2m_1} \left[
  \frac{\partial^2 b}{\partial x^2} + 2 \frac{\partial b}{\partial x}
  \frac{ (\partial \psi_1 / \partial x)}{\psi_1} \right] 
+ \frac{i \hbar}{2m_2} \left[ d - b^2 \right] +  \frac{dX_2}{dt}
\left[ c - a b\right] \nonumber \\
&& \; \; \; - \frac{2i}{\hbar} a \frac{\partial V}{\partial x_2} \Big|_{x_2 =
  X_2(t)}  - \frac{i}{\hbar} \frac{\partial^2 V}{\partial x_2 ^2} \Big|_{x_2 = X_2(t)}.
\label{eq-bde}
\end{eqnarray}
The $c$ and $d$ which appear here need their own time-evolution
equations (which will in turn involve further potentials $e$ and $f$),
and so on.   The result is a countably infinite network of potential fields
obeying coupled time-evolution equations.  These potentials then of
course appear in the Schr\"odinger-type equations governing the
single-particle pilot-wave fields which guide the particles.  The
exact statistical predictions of quantum theory are reproduced, but
the configuration space wave function $\Psi$ is nowhere to be found.
We used $\Psi$, of course, to find out how the potentials $a$, $b$, $c$,
etc., must interact and evolve in order to reproduce the usual Bohmian
particle trajectories.  But once we have Equations \eref{eq-v1eff2} as
well as \eref{eq-ade}, \eqref{eq-bde}, etc., the universal wave
function $\Psi$ can, like the proverbial ladder, be simply kicked away.

Thus, the answer to the question posted in the paper's title is:  yes.
It \emph{is} possible to reproduce the exact particle trajectories of Bohmian
mechanics, and hence the exact statistical predictions of ordinary
quantum mechanics, in a theory in which the configuration space wave
function $\Psi$ is replaced with single-particle wave functions (one
for each particle) in ordinary physical space.  One merely needs to
introduce appropriate one-particle effective potentials.  

But the price of defining these potentials in the way we have done here -- with
an \emph{infinite} number of interacting potential fields associated
with each particle -- seems rather high.  We thus turn to attempting
to develop an alternative, less ontologically complex, approach to defining appropriate
one-particle potentials.

\section{An empirically viable theory in 3d physical space with modest ontological complexity}
\label{sec4}

It is not out of the question that some radical new perspective on the
problem will allow, in the future, a completely simple and
straightforward definition of the one-particle effective potentials
needed to reproduce the quantum predictions in the context of a
Bohm-inspired theory of particles being guided by one-particle
pilot-waves in physical space.  Such an innovation might be analogized
to the simplification that was afforded in the description of
planetary orbits when Kepler abandoned the axiom of explaining the
orbits in terms of exclusively circular motion.  

Unfortunately, such a development remains a speculative fantasy.  But
we can develop a more realistic, if also more mundane, alternative
definition for the one-particle effective potentials by considering
an ontological simplification arrived at by approximating
the infinite network introduced in the previous Section.  
After all, as illustrated by the dynamical collapse (GRW) type theories \citep{GRW1,GRW2},
perfect agreement with quantum predictions is not sacrosanct.  What
matters is instead agreement with experimental data.  Disagreements
with the quantum predictions are perfectly tolerable if they can be
confined to situations where no experimental data is yet available,
and such options are particularly welcome if they help to address
foundational and/or philosophical problems.  

That, then, indicates the nature of our proposal:  if the price of
eliminating the configuration space wave function $\Psi$ is the
introduction of an infinitely complex network of interacting
single-particle potential fields, it is not clear that much is
gained.  However, if the network of single-particle potential fields
can be made reasonably simple, while still maintaining empirical
adequacy, this would be a very compelling argument that the
configuration space wave function (and hence the torturous
philosophical conundrums its presence raises) might be simply
eliminated.   In the remainder of this Section we will present
preliminary evidence in support of this possibility, starting with
another look at the case of non-entangled particles.

\subsection{ The case of no entanglement }
\label{separable} 

We argued in \sref{non-entangled} that when the universal wave function factorizes,
the two conditional wave functions are given by $\alpha$ and $\beta$, respectively, and these
should obey the expected one-particle Schr\"odinger equations.  Let us
now see how this same conclusion arises from the general
Schr\"odinger-type equation governing the time evolution of the conditional wave functions.  

The crucial point is that 
with $\Psi(x_1,x_2,t) = \alpha(x_1,t) \beta(x_2,t)$, we have 
\begin{equation}
\psi'_1(x,t) = \alpha(x,t) \frac{\partial \beta}{\partial x_2}
\Big|_{x_2 = X_2(t)}
\end{equation}
so that 
\begin{equation}
A_1(t) = i \hbar \frac{d X_2}{dt} \frac{ \partial \beta / \partial
  x_2}{\beta} \Big|_{x_2 = X_2(t)}
\end{equation}
is \emph{independent of} $x$.  (This happened because, when there is
no entanglement, the strange object $\psi'_1(x,t)$ is
\emph{proportional to} the conditional wave function, $\psi_1(x,t)$.  And so the
potential $A_1$, which depends on the \emph{ratio}, has no
$x$-dependence.)  

Similarly,
\begin{equation}
\psi''_1(x,t) = \alpha(x,t) \frac{\partial^2 \beta}{\partial x_2^2}
\Big|_{x_2 = X_2(t)}
\end{equation}
so that
\begin{equation}
B_1(t) = - \frac{\hbar^2}{2m_2} \frac{ \partial^2 \beta / \partial
  x_2^2}{\beta} \Big|_{x_2 = X_2(t)}
\end{equation}
is also \emph{independent of} $x$.  

Thus, whenever two particles are
un-entangled, the only $x$-dependence in the effective potential
$V_1^{\text{eff}}$ arises from the conditional potential,
$V[x,X_2(t),t]$.  And of course, for non-interacting particles, this
conditional potential is simply $V_1(x,t)$ (the external potential experienced by
particle 1) plus a (perhaps time-dependent) constant,
$V_2(X_2(t),t)$, which simply introduces an uninteresting overall
time-dependent phase into $\psi_1(x,t)$.  

The important point here is that the mysterious potential energy terms,
$A_1$ and $B_1$, do not depend on $x$ and hence also have no
meaningful influence on $\psi_1(x,t)$, so long as the two-particle
state remains unentangled.  

\subsection{ The small entanglement approximation}      

We saw in the last subsection that, as long as our two particles remain
unentangled, the  terms $A_1$ and $B_1$ are functions
of $t$ only and hence affect the single-particle wave function only by
giving it an overall time-dependent multiplicative factor which
cancels out in the guidance formula, Equation \eqref{eq-guidance2}.
The terms $A_1$ and $B_1$, in other words, might as well be set to
zero.
In this situation, the meaningful and important contribution to
$V_1^{\text{eff}}$ comes exclusively from the conditional potential,
$V[x,X_2(t),t]$.  

This suggests a kind of first-order approximation to the infinitely
complex network of single-particle potentials developed at the end of
the previous Section:  we retain, for each particle, the conditional potential term in
$V_i^{\text{eff}}$ and set the other terms to zero:
\begin{equation}
V_i^{\text{eff}} \approx V(x_1, x_2, t) \Big|_{x_i = x \text{ and }
  x_n = X_n(t) \forall n \ne i}.
\end{equation}
On the grounds that
this should work perfectly when there is no entanglement, we call this
the ``small entanglement approximation'' or SEA.  But note that there is no
clear \emph{a priori} reason to expect the SEA to work well when there is
even just a little entanglement.  It could be, for example, that as
soon as any entanglement develops, the small entanglement
approximation breaks down completely, giving particle trajectories 
that are wildly unphysical and/or blatantly different from those of
ordinary Bohmian mechanics.  

In principle, a gifted theoretician could perhaps intuit or calculate
the accuracy of the small entanglement approximation in various kinds
of situations.  Unfortunately, the present authors lack the requisite
intuition.  So we have turned instead to solving the relevant
equations numerically, and comparing the results of the small
entanglement approximation to the results obtained by solving the
configuration space Schr\"odinger equation for $\Psi$.  

In the next subsection we present a numerical example of two particles
continuously interacting through a non-separable potential.

\subsection{Numerical example for two interacting particles}

In the following example, we show a scenario where the \emph{small entanglement approximation} 
works quite well for a system of continuously interacting
particles. In particular, we consider two
particles, each moving in the presence of a harmonic external
potential, and also interacting:
\begin{equation}
V(x_1,x_2) = F \left(x_1^2+x_2^2\right) + C x_1x_2
\label{eq-potential}
\end{equation}
with $F=10^{12}$ $eV \cdot m^{-2}$.  Since we consider an initial product state wave function $\Psi(x_1,x_2,0) = \alpha(x_1)\beta(x_2)$,
entanglement will develop  as a result of the
interaction term $Cx_1 x_2$. The
parameter $C$ may therefore be used to quantify the \emph{amount of entanglement}
between the particles.  We anticipate that the small entanglement
approximation will be increasingly accurate for smaller values of
$C$. 

According to the SEA, the Schr\"odinger equation satisfied by the
single-particle wave function $\psi_1(x,t)$ of particle 1 is
\begin{eqnarray}
i \hbar \frac{\partial \psi_1(x,t)}{\partial t} &=& -
\frac{\hbar^2}{2m_1}\frac{\partial^2 \psi_1(x,t)}{\partial x^2} +
V[x,X_2(t)] \psi_1(x,t) \label{eq-particle1} \end{eqnarray}
while the corresponding equation for 
$\psi_2(x,t)$ is
\begin{eqnarray}
i \hbar \frac{\partial \psi_2(x,t)}{\partial t} &=& -
\frac{\hbar^2}{2m_2}\frac{\partial^2 \psi_2(x,t)}{\partial x^2} +
V[X_1(t),x] \psi_2(x,t) \label{eq-particle2}.
\end{eqnarray}
It is very important to realize that Equations \eref{eq-particle1} and
\eref{eq-particle2} are coupled through the Bohmian trajectories. This
coupling is responsible for the interaction between the two degrees of
freedom. The trajectories $X_1(t)$ and $X_2(t)$ are computed from the
associated conditional wave functions at each time step.  There is no
need to track (or even mention) the configuration
space wave function $\Psi(x_1,x_2)$.  

\begin{figure*}
\centering
\includegraphics[width=0.75\textwidth]{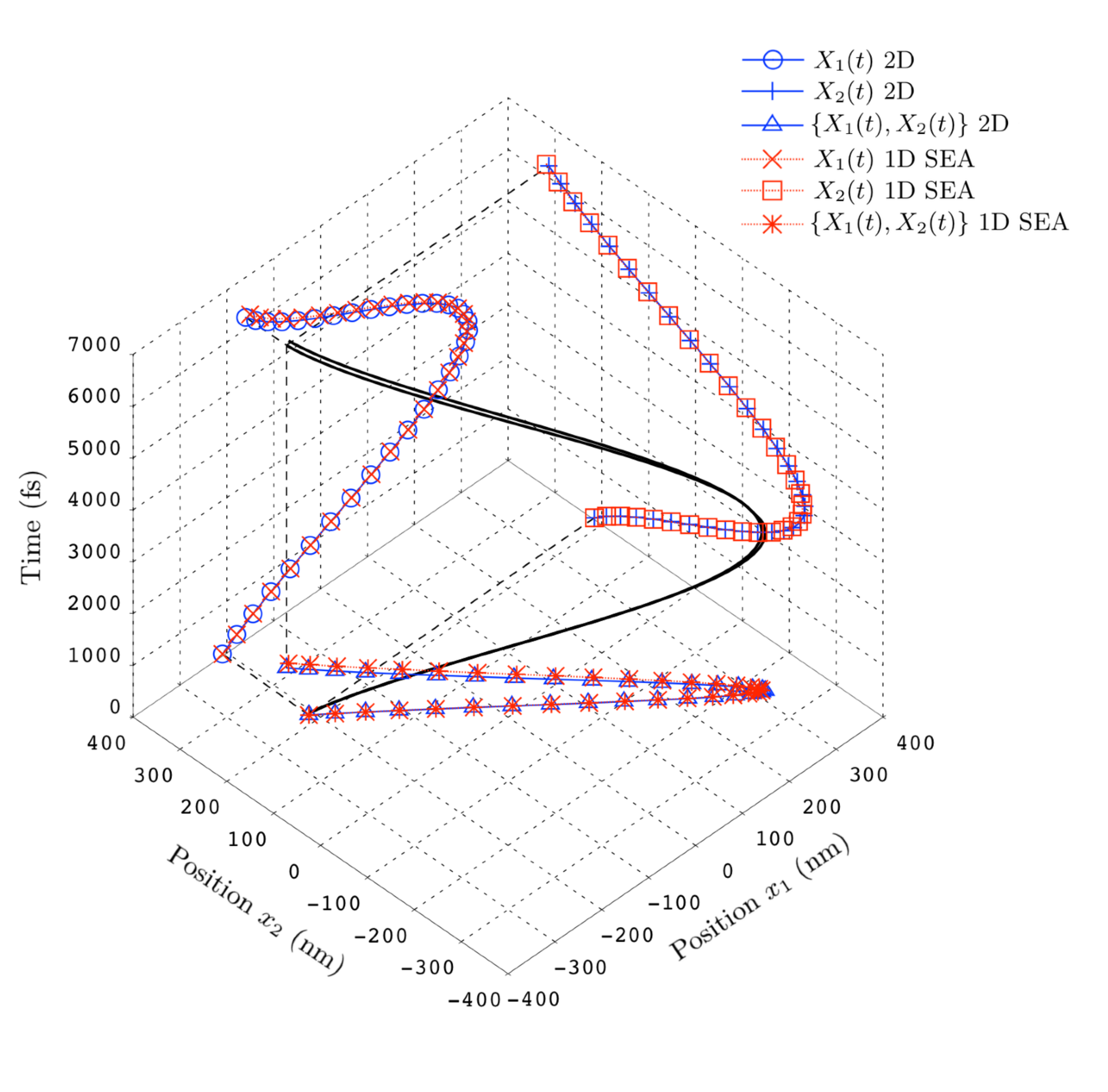}
\caption{Comparison between the Bohmian trajectories calculated from the 2D solution (with $\odot$, $+$ and $\triangle$ in blue) of the Schr\"odinger equation (Equation \eqref{eq-2dexact}) and the 1D solution (with $\times$, $\square$ and $*$ in red) of the small entanglement approximation (Equations \eqref{eq-particle1}-\eqref{eq-particle2}) for $C=-1~\cdot~10^{12}$~$eV \cdot m^{-2}$. We also plot with black solid lines the trajectories in the configuration space for both solutions}
\label{conditional wave function-figure3}
\end{figure*}

In order to check the accuracy of the small entanglement
approximation, however, the results obtained from
Equations \eqref{eq-particle1}-\eqref{eq-particle2} will be compared with
the solution of the following two-particle Schr\"odinger
equation:
\begin{equation}
i \hbar \frac{\partial \Psi(x_1,x_2,t)}{\partial t} = \left[ - \frac{\hbar^2}{2m_1}\frac{\partial^2}{\partial x_1^2} - \frac{\hbar^2}{2m_2}\frac{\partial^2}{\partial x_2^2} + V(x_1,x_2) \right] \Psi(x_1,x_2,t).
\label{eq-2dexact}
\end{equation}
We consider $m_1=m_2=m_0$ with $m_0$ the free electron mass. In
\fref{conditional wave function-figure3} we plot the trajectories
computed from the 2D solution of the Schr\"odinger equation
\eref{eq-2dexact}  and from the 1D small entanglement approximation,
Equations \eqref{eq-particle1}-\eqref{eq-particle2}. 
In this particular example, we use $C=-1 \cdot 10^{12}$ $eV \cdot
m^{-2}$ in Equation \eqref{eq-potential}. We see that there are no significant
differences between the 2D and 1D computation of
the trajectories $X_1(t)$ and $X_2(t)$. 

\begin{figure*}
\centering
\includegraphics[width=0.75\textwidth]{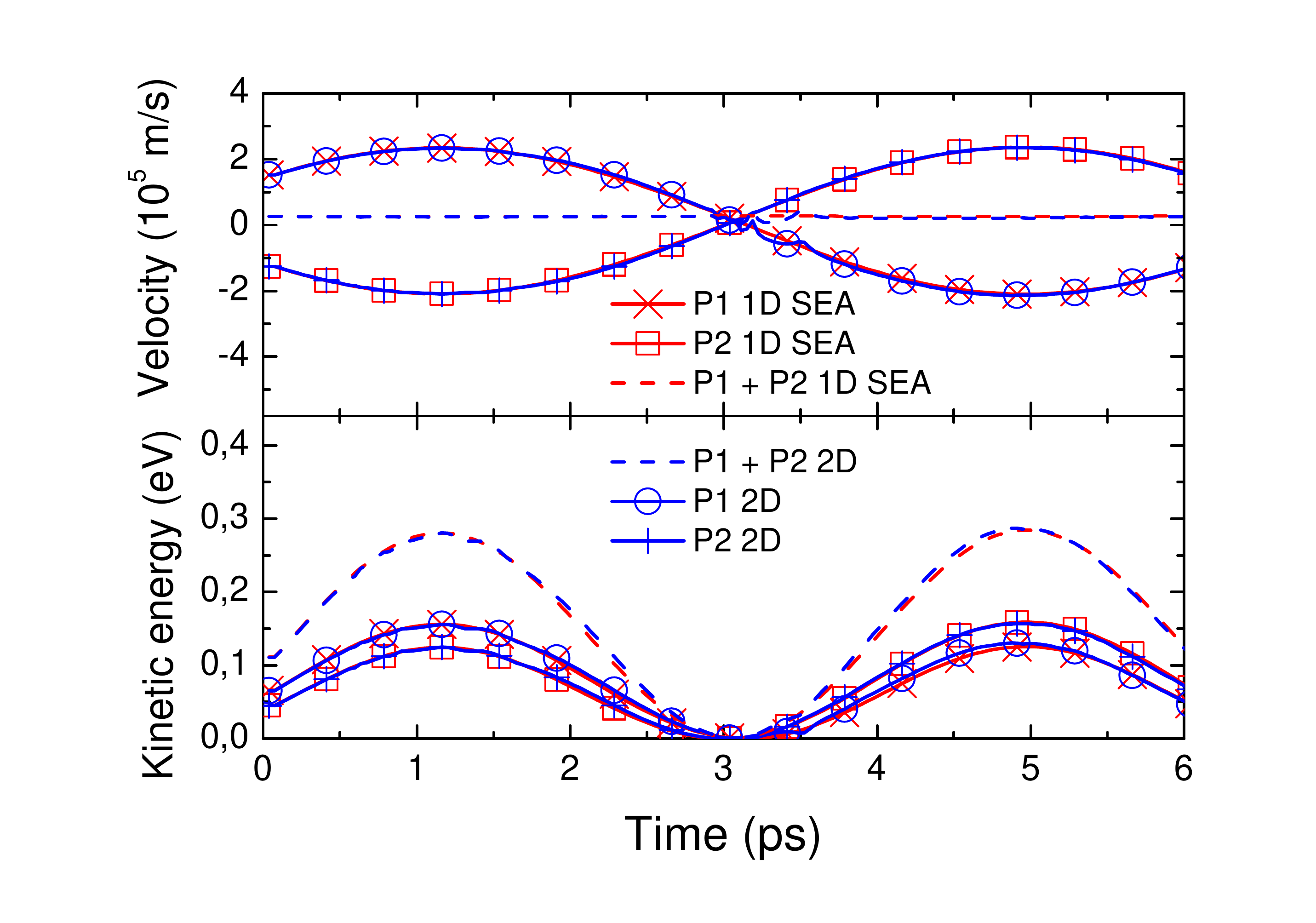}
\caption{Bohmian velocity and kinetic energy for particle 1 (P1) and particle 2 (P2) as function of time for $C=-1~\cdot~10^{12}$~$eV \cdot m^{-2}$. Dashed lines represent the sum of the velocity and the kinetic energy for particle 1 plus particle  2 (P1 + P2)}
\label{conditional wave function-figure4}
\end{figure*}

In \fref{conditional wave function-figure4} we plot the Bohmian
velocity and the kinetic energy for the two particles comparing the
standard Bohmian mechanics solution based on $\Psi$ (``2D'')  with the small
entanglement approximation to the theory developed here (``1D'').
No significant differences are present. We emphasize that there is an interchange of energy between the first and second particle, showing the non-separable quantum nature of the studied system. 

We repeat the same analysis with a different value of the interaction
parameter $C=-2~\cdot~10^{12}$~$eV \cdot m^{-2}$. In \fref{conditional
  wave function-figure5}, we see that the difference between the two
cases are quite small but detectable. 
\begin{figure*}
\centering
\includegraphics[width=0.75\textwidth]{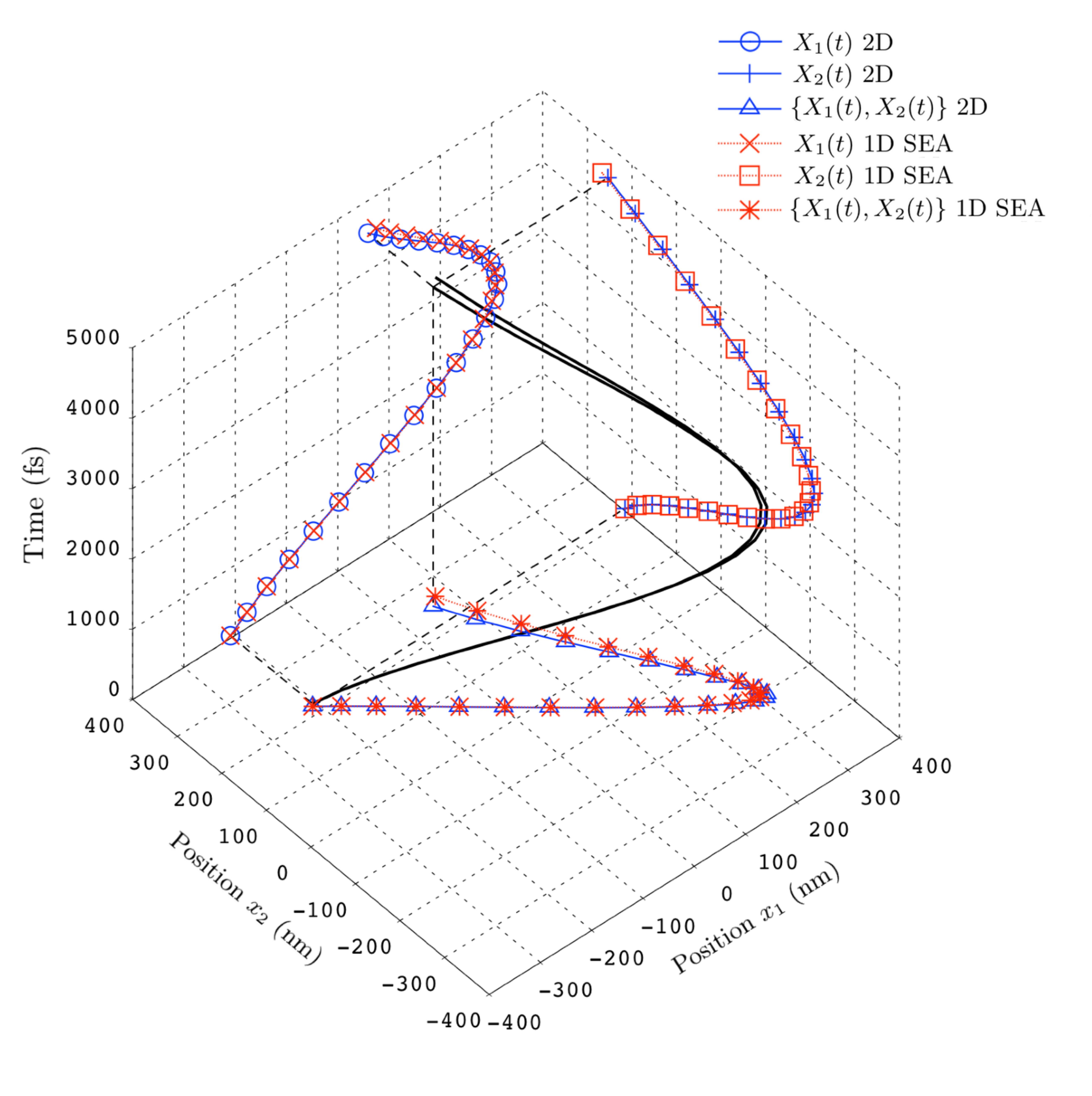}
\caption{Comparison between the trajectories calculated from the 2D solution of the Schr\"odinger equation (Equation \eqref{eq-2dexact}) and the 1D solution of the small entanglement approximation (Equations \eqref{eq-particle1}-\eqref{eq-particle2}) for $C=-2~\cdot~10^{12}$~$eV \cdot m^{-2}$. Symbols are the same as in \fref{conditional wave function-figure3}}
\label{conditional wave function-figure5}
\end{figure*}
In order to confirm this fact we report in \fref{conditional wave function-figure6} the velocity and kinetic energy for particle 1 and 2. We clearly see a difference in the two solutions. 
\begin{figure*}
\centering
\includegraphics[width=0.75\textwidth]{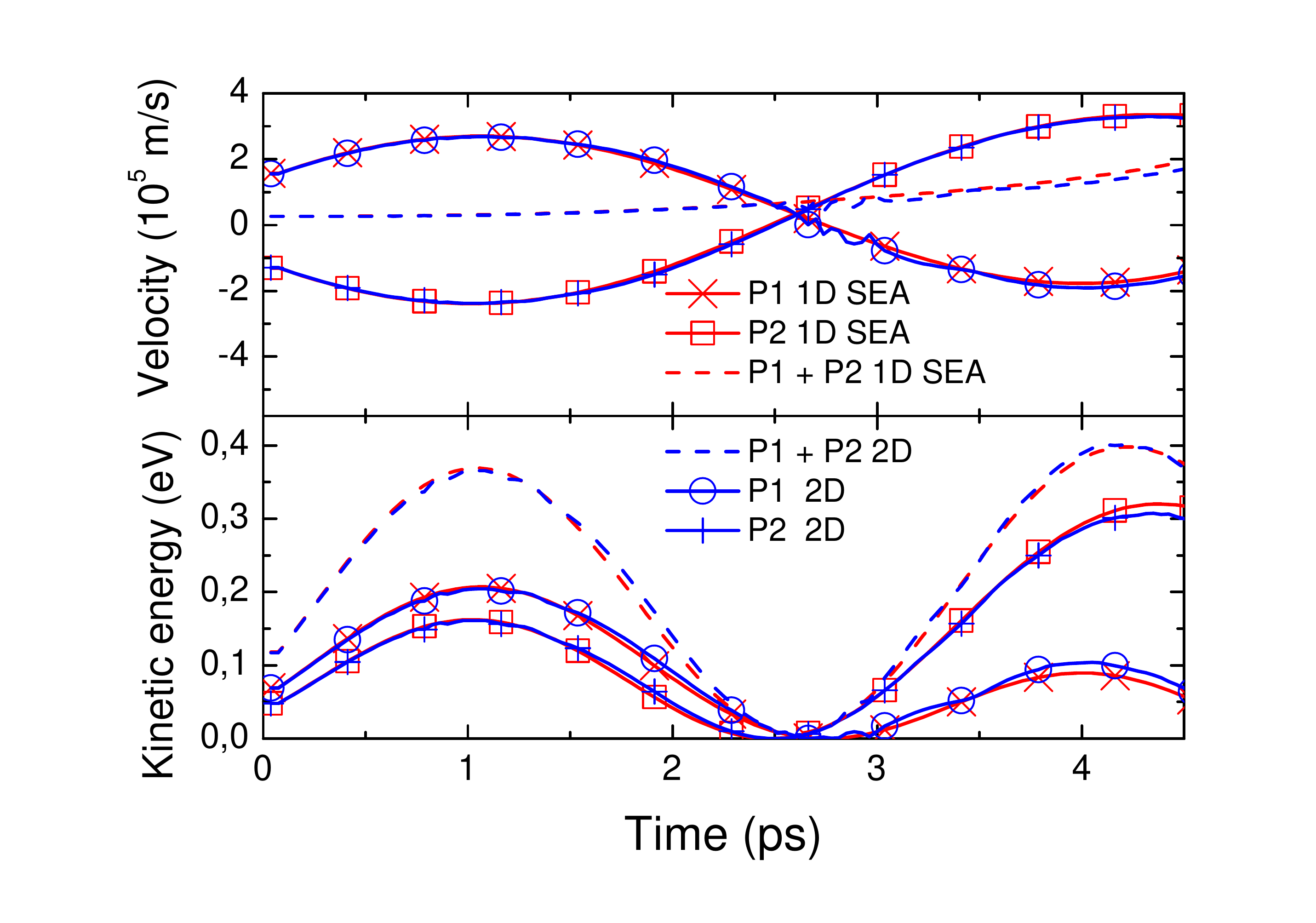}
\caption{Bohmian velocity and kinetic energy for the two particles in function of time for $C=-2~\cdot~10^{12}$~$eV \cdot m^{-2}$. Symbols are the same as in \fref{conditional wave function-figure4}}
\label{conditional wave function-figure6}
\end{figure*}

\begin{figure*}
\centering
\includegraphics[width=0.75\textwidth]{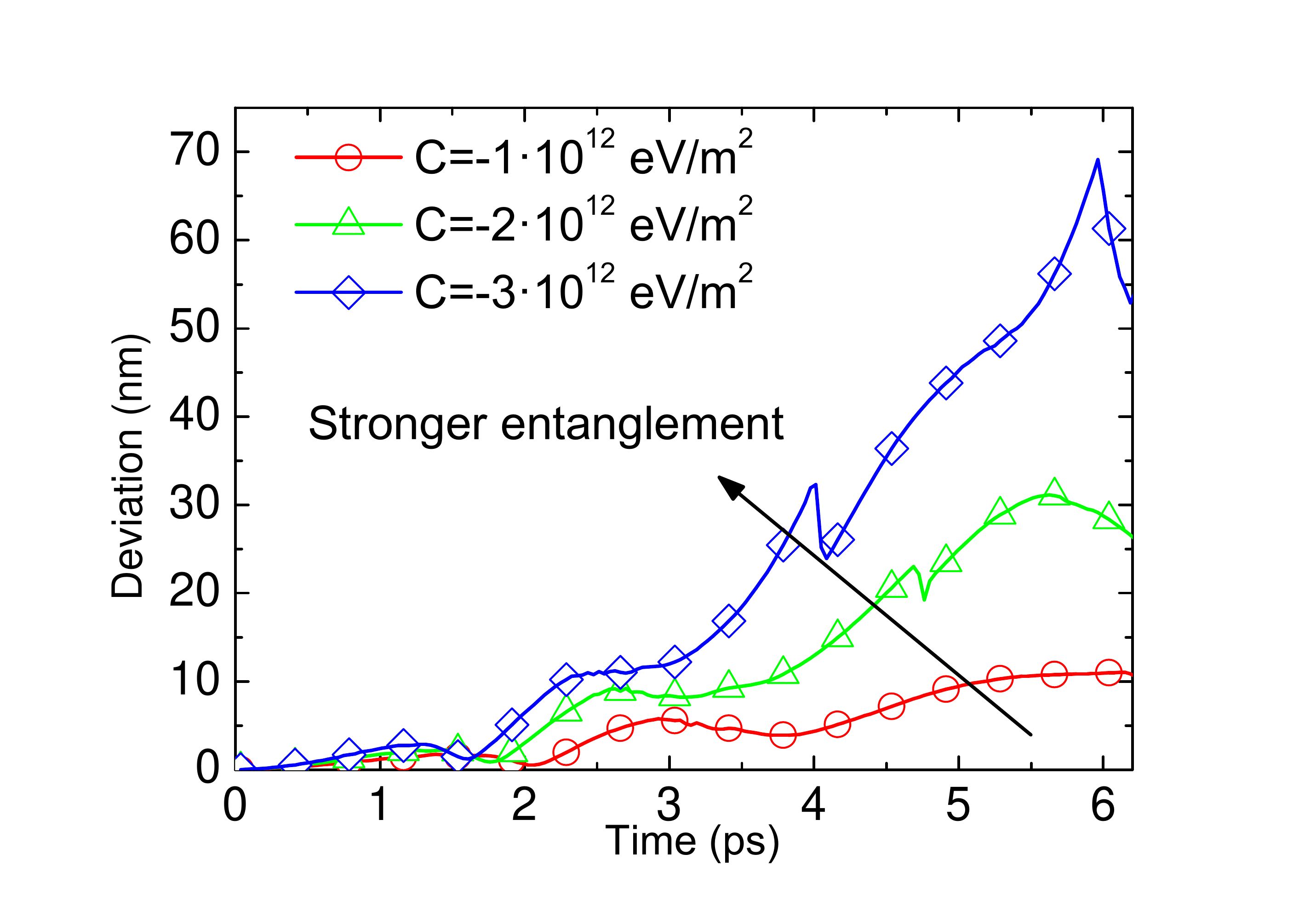}
\caption{Deviation computed from Equation \eqref{eq-error} as function of time for different values of $C$}
\label{conditional wave function-figure7}
\end{figure*}

In order to show quantitatively what is the deviation when the entanglement between the two particles grows, we compute the absolute error of the 2D, $X^{\text{2D}}_i(t)$, and the 1D,  $X^{\text{1D}}_i(t)$, trajectories as:
\begin{equation}
\text{Deviation}(t) = \sqrt{\left(X^{\text{2D}}_1(t) - X^{\text{1D}}_1(t)\right)^2+ \left(X^{\text{2D}}_2(t) - X^{\text{1D}}_2(t)\right)^2}.
\label{eq-error}
\end{equation}
In \fref{conditional wave function-figure7} we plot
Equation \eqref{eq-error} for three value of $C$. First, we see that the
deviation increases as time grows. Secondly, the deviation increase for larger
absolute values of $C$ confirming that our SEA is increasingly
accurate for smaller entanglements.

We conclude that the computation of one-particle wave functions from
Equation \eref{equation} with the small entanglement approximation works quite well in the scenario
governed by Equation \eref{eq-potential}. 
This type of computation has already been used to study
quantum transport in nano electronic devices \citep{albareda,
  OriolsBook,albaredaPRB1,albaredaPRB2,Fabio,alarcon}. A commercial
software named BITLLES\footnote{ See the website \emph{http://europe.uab.es/bitlles}} has been developed following
these ideas.
 Finally, to be fair, let us mention that there are many
other shapes of potentials where this simplest approximation for the
conditional wave function does not work as well. In any case, we
emphasize that the approximation just presented is only the first step
(although quite accurate and sufficient in some cases) towards a
method able to tackle quantum phenomena with appreciable entanglement using
only one-particle wave functions in the physical 1D (or 3D) space.

The previous computations in the small entanglement approximation
(SEA) strongly suggest that the type of theory sketched at the
beginning of this Section can be made empirically viable. As already noted,  perfect
agreement with the quantum predictions is not required if the
disagreement is limited to cases where no experimental data is yet available. 
The fact that the most brute approximation (what we call the SEA)
already works quite well in some situations provides strong evidence
that a more seriously realistic approximation (in which the network of
potentials is cut, for example, between $b$ and $c$) could adequately reproduce
the effects (such as non-locality and non-unitarity) of the full effective potentials.  

The question of where, exactly, the tower of potentials should be cut
is the subject of ongoing work by the authors.  To be clear, though,
this is not something that would be adjusted on a case-by-case basis
for purely calculational purposes.  Instead the idea is to propose 
that the ontology involves some
large number $N$ of sets composed of:  a point particle, an associated
pilot-wave field, and -- say -- 5 dynamical potential fields ($a$,
$b$, $c$, $d$, and $e$).  The choice of where to cut the tower, that
is, is like the choice of values for the parameters (which define the
length scale and frequency of the spontaneous collapses) in GRW-type
theories.  It is made once and for all, in order to try to achieve
empirical adequacy, and then the theory says what it says. What it says will include some empirically testable deviations from the predictions of ordinary quantum mechanics, in situations (for example) involving spatially and/or temporally long-range entanglement.

\section{Summary and Future Prospects}
\label{sec5}

The fundamental lesson of our paper is that it is possible to describe quantum
phenomena with a set of single-particle pilot-wave fields in ordinary
3-dimensional physical space instead of the usual
(ontologically-puzzling) configuration space wave function.  Here we
summarize the main sub-points:

\begin{itemize}

\item \emph{The conditional wave function:} The ordinary quantum theory offers no way to define the wave function for a single part of a larger quantum mechanical system, and it offers
only the notoriously vague and problematic measurement axioms to
describe how quantum mechanical systems interact with measuring
instruments. In Section \ref{sec2}, we briefly reviewed Bohmian mechanics showing how it provides a natural notion of the (conditional) wave function of a subsystem \citep*{bm,bm2} and allows the usual quantum measurement formalism (including for example the collapse postulate) to be \emph{derived} from the fundamental dynamical postulates of the Bohmian theory. The (Bohmian) conditional wave function offers a very relevant and unique starting point to formulate quantum mechanics in terms of single-particle wave functions in a 3-dimensional  physical space.

\item  \emph{Single-particle wave functions with an infinite network of interacting potentials fields:} In Section \ref{sec3}, we have shown that it is indeed possible to formulate an empirically
adequate Bohmian-type quantum theory in which the usual 
wave function $\Psi$ (the wave function of the
universe, living in $3N$-dimensional configuration space) is replaced
by $N$ single-particle wave functions that live in $3$-dimensional
physical space.  In particular, we have shown explicitly 
how to define a (countably
infinite) network of interacting potential fields which (together
with the usual classical potential function $V$) drive the
single-particle wave functions in such a way that the particle
trajectories exactly reproduce those of standard Bohmian mechanics.
The exact empirical predictions of quantum mechanics (including
phenomena involving entanglement and non-locality) can therefore in
principle be reproduced by a theory in which the
philosophically-puzzling wave function $\Psi$ plays no role. This result, reformulated here in terms of interacting potentials fields, was already anticipated in \cite{telb}. We emphasize that the \emph{non-locality} required by Bell's theorem can actually be embedded in a theory of exclusively \emph{local} beables.

This result may initially seem surprising in light of the recent
``PBR'' theorem of \citep{pbr}, according to which the wave
function $\Psi$ must be regarded as physically real in any theory that
shares the exact predictions of quantum mechanics.  The
main point is that ``physically real'' (for PBR) means that the mathematical object in question is a
function of the ontic state posited by the theory.  For
the kind of theory proposed here, the full ontic state includes not
only the particle positions and the states of each single-particle
pilot-wave field, but also the network of interacting single-particle
potentials.  We have shown explicitly how the single-particle
potentials can be defined in terms of the universal wave function
$\Psi$; it is hardly surprising, then, that $\Psi$ is determined by
the complete ontic state.\footnote{In fact, $\Psi$ is determined by
  the one-particle wave function and the associated potentials
  \emph{for any one particle}.  The \emph{complete} ontic state --
  comprising the one-particle wave function and associated potentials
  for \emph{all} particles, plus the particle positions themselves -- thus
  contains a tremendous amount of redundancy.  This is yet another
  strong piece of evidence suggesting that empirical viability should
  be able to be achieved, even with a greatly reduced ontic
  complexity.}   
Thus, for the type of theory proposed here, $\Psi$ is indeed
``physically real'' in the sense of \citep{pbr}.  Superficial
appearances to the contrary notwithstanding, however, this is perfectly
compatible with our main point, namely, that one need not regard $\Psi$ as
``physically real'' in the straightforward, direct sense of ``View 1''
and ``View 2'' from the Introduction.   We have shown, instead, how
$\Psi$ can instead be viewed as an indirect and abstract
characterization of the state of a certain constellation of
interacting \emph{local beables}.  That is, we have shown how $\Psi$
can be regarded as having the same (totally unproblematic) status that
is usually assigned to, say, Hamilton's principal function $S(x_1,
x_2, ..., x_N, t)$ in classical mechanics.

\item \emph{Single-particle wave functions with reduced ontological complexity:}
In Section \ref{sec4}, in addition to the demonstrations of the plausibility of an explanation of the quantum theory with local beables, we also took a
preliminary step toward an ontologically simpler theory (of the
general sort proposed above), in which the infinite network of
interacting potentials is reduced to a more reasonable size at the
price of introducing some disagreements with the predictions of
ordinary quantum theory.  We showed explicitly that, even making the
most draconian imaginable cuts to the network, one still gets
reasonable behavior and indeed rather surprisingly good agreement with
the predictions of quantum mechanics in certain simple situations.  
Of course, certain important
effects like non-locality and non-unitary evolution (``collapse'')  
cannot be reproduced in
this so-called ``small entanglement approximation''.  But our results
provide a basis for optimism that such phenomena  (and ultimately all
currently available empirical data) might be reproduced in a theory
that goes beyond the SEA, i.e., a theory with a moderate and tolerable
degree of ontological complexity.  Such a theory would then have a
status comparable to that already enjoyed by GRW-type theories:  it
would be empirically viable despite making predictions that are (in
certain ``exotic,'' as-yet-untested cases) different from those of
ordinary quantum mechanics.  (And note that, as soon as deviations
from quantum predictions are contemplated, the PBR theorem will no
longer entail that $\Psi$ be ``physically real'', even in PBR's
somewhat misleading
sense.  So there is no contradiction in the idea of constructing an
empirically viable theory using some restricted subset of the network
of potential fields, even though $\Psi$ itself could not be computed from
the single-particle wave functions and the restricted subset.)

\end{itemize}

While we have shown in principle that the wave function in
configuration space \emph{can} be replaced by single-particle wave
functions in physical space, the details of how best to accomplish
``View 3'' remain unsettled.  Ongoing work by the authors will explore
the effects of cutting the network of potentials between $b$ and
$c$ --instead of, as was done here in our preliminary SEA, cutting it
before $a$. This should be particularly interesting in that certain
key effects of the full effective potential (such as non-locality and
non-unitarity) should be included.   

But in general ``View 3'' remains a young and as-yet unproven research
program.  Whether it will bear important fruit remains to be seen.
But it should already be clear that it is simply premature to debate,
for example, whether the universal wave function $\Psi$ is better
regarded as a physically-real field in a physically-real configuration
space or instead as a new and unusual (perhaps time-dependent) type
of natural law.  \emph{There are other possibilities.}  In particular,
one alternative possibility is the one 
suggested already by certain abstract reformulations
of classical mechanics, in which the unusual time-dependent function
on configuration space is regarded as an indirect description of some
more mundane and more familiar and less puzzling pieces of physical
ontology.  

Let us close by returning to the historical considerations with which
we began in the introduction.  Shortly after inventing wave mechanics and his eponymous
equation, Schr\"odinger noted that ``the use of the [configuration]
space is to be seen only as a mathematical tool, as it is often
applied also in the old mechanics; ultimately ... the process to be
described is one in space and time.''  \citep[p. 447]{bv}
This overall sentiment -- that the wave function in configuration
space cannot be taken seriously as corresponding directly to a
physically-real wave in a physically-real, high-dimensional space -- 
seems to have been shared by most of the
prominent realist-leaning physicists at the time:  not just
Schr\"odinger, but (as we saw in the Introduction) Einstein, and also
de Broglie, who explained in his address at the 1927 Solvay conference
that ``if one wants to \emph{physically} represent the evolution of a
system of corpuscles, one must consider the propagation of $N$ waves
in space...''  \citep[p. 79]{bv}
Indeed, as summarized by Linda Wessels based on a 1962 interview with
Carl Eckart (conducted by John Heilbron), it would be desirable ``to
rewrite the equations of wave mechanics so that even for a system of
several `particles', only 3-dimensional wave functions would be
determined.  C. Eckart has reported that at one time he attempted this
and remarked that it was something that initially `everybody' was
trying to do.''  \citep{wessels}

Of course, eventually the more positivist/instrumentalist outlook of
Bohr and Heisenberg won out, and this concern about physically
interpreting the nature of the quantum wave function largely died
away.  It is a very positive development that, in recent decades, the
shortcomings of the positivist approach are being increasingly
recognized and interpretational questions are again being taken
seriously by physicists and philosophers.  Our main message can then
be summarized as follows:  as we re-open the discussion about the
physical meaning and significance of the quantum mechanical wave
function, let us not forget about the original view of Einstein,
Schr\"odinger, de Broglie, Eckart, and others -- that the
configuration space wave function must be some kind of abstract,
indirect description of physical processes in ordinary physical
space.  As we have shown, this prematurely-abandoned program remains
viable, and the pilot-wave theory of de Broglie and Bohm (with its
natural way of defining one-particle wave functions in physical space)
provides an especially promising starting point.  It may yet vindicate
Schr\"odinger, who expressed the hope, already in 1927, that
``in the end everything will indeed become intelligible in three
dimensions again.''  \citep[p. 461]{bv}

\begin{acknowledgements}
We gratefully acknowledge Nino Zangh\`{i} and Albert Sol\'{e} for reading a preliminary version of this paper and for very useful discussions. D.M. is supported in part by INFN and acknowledges support of COST action (MP1006) through STSM. X.O. acknowledge support from the \lq\lq{}Ministerio de Ciencia e Innovaci\'{o}n\rq\rq{} through the Spanish Project TEC2012-31330 and by the Grant agreement no: 604391 of the Flagship initiative  \lq\lq{}Graphene-Based Revolutions in ICT and Beyond\rq\rq{}.
\end{acknowledgements}

\end{document}